\def\e{{\rm e}}
\def\del{\partial}
\def\half{{1\over2}}

\def\abs#1{{\left|{#1}\right|}}
\def\vev#1{\langle #1 \rangle}
\def\del{\partial}
\def\dslash{\del\kern-0.55em\raise 0.14ex\hbox{/}}

\def\rough#1{\raise.3ex\hbox{$#1$\kern-.75em\lower1ex\hbox{$\sim$}}}
\newcommand{\PRD}[3]{{\it Phys. Rev.} {\bf D{#1}} (19{#3}) {#2}}
\newcommand{\PRDM}[3]{{\it Phys. Rev.} {\bf D{#1}} (20{#3}) {#2}}

\newcommand{\NPB}[3]{{\it Nucl. Phys.} {\bf B{#1}} {#2} (19{#3})}
\newcommand{\NPBM}[3]{{\it Nucl. Phys.} {\bf B{#1}} (20{#2}) {#3}}
\newcommand{\PLB}[3]{{\it Phys. Lett.} {\bf {#1}B} (19{#3}) {#2}}
\newcommand{\PLBM}[3]{{\it Phys. Lett.} {\bf {#1}B} (20{#3}) {#2}}

\newcommand{\ANN}[3]{{\it Ann. Phys. (N.Y.)} {\bf {#1}}, {#2} (19{#3})}
\newcommand{\PR}[3]{{\it Phys. Rep.} {\bf {#1}}, {#2} (19{#3})}

\newcommand{\MPL}[3]{{\it Mod. Phys. Lett.} {\bf A{#1}} (19{#3}) {#2}}
\newcommand{\MPLM}[3]{{\it Mod. Phys. Lett.} {\bf A{#1}} (20{#3}) {#2}}

\newcommand{\jhep}[3]{{\it J. High Energy Phys.}{\bf {#1}}, {#2} (20{#3})}
\newcommand{\hepth}[1]{{\tt [hep-th/#1]}}
\newcommand{\hepph}[1]{{\tt [hep-ph/#1]}}
\newcommand{\hmu}{\hat\mu}
\newcommand{\hnu}{\hat\nu}

\documentclass[12pt]{article}
\usepackage{graphicx}
\begin{document}
\baselineskip=18pt
%%%%%%%%%%%%%%%%%%%%%%%%%%%%
\begin{titlepage}
%%%%% PREPRINT NUMBERS %%%%%%
\begin{flushright}
KUNS-1942\\
OU-HET-498/2004
%hep-th/0305xxx
\end{flushright}
%%%%%%%%%%%%%%%%%%% TITLE %%%%%%%%%%%%%%%%%%
%\begin{center}{\Large\bf Correct Effective Potential 
%for Broken Supersymmetric Yang-Mills 
%Theory on $M^4\times S^1$}
%\end{center}
%%%%%%
\begin{center}{\Large\bf Correct Effective Potential 
of Supersymmetric Yang-Mills 
Theory on $M^4\times S^1$}
\end{center}
%%%%%%%%%%%%%%%% AUTHORS %%%%%%%%%%%%%%%%%%%%%%%
\vspace{1cm}
\begin{center}
Naoyuki {Haba}$^{(a),}$
\footnote{E-mail: haba@ias.tokushima-u.ac.jp}
Kazunori {Takenaga}$^{(b),}$
\footnote{E-mail: takenaga@het.phys.sci.osaka-u.ac.jp}
Toshifumi {Yamashita}$^{(c),}$
\footnote{E-mail: yamasita@gauge.scphys.kyoto-u.ac.jp},
\end{center}
%%%%%%%%%%%%%%%%%%%%%%% AFFILIATION %%%%%%%%%%%%
\vspace{0.2cm}
\begin{center}
%\small
${}^{(a)}$ {\it Institute of Theoretical Physics, University of
Tokushima, Tokushima 770-8502, Japan}
\\[0.2cm]
${}^{(b)}$ {\it Department of Physics, Osaka University, 
Toyonaka, Osaka 560-0043, Japan}
\\[0.2cm]
${}^{(c)}${\it Department of Physics, Kyoto University, Kyoto, 
606-8502, Japan}
%%%%%
%${}^{(d)}$ {\it School of Theoretical Physics,
%Dublin Institute for Advanced Studies, \\
%10 Burlington Road, Dublin 4, Ireland}
%%%%%%%
\end{center}
%%%%%%%%%%%%%%%%%% ABSTRACT %%%%%%%%%%%%%%%
\vspace{1cm}
\begin{abstract}
We study an ${\cal N}=1$ supersymmetric Yang-Mills theory 
defined on $M^4\times S^1$. The vacuum 
expectation values for adjoint scalar field 
in vector multiplet, though important, has been overlooked in 
evaluating one-loop effective potential of the 
theory. We correctly take the vacuum
expectation values into account in addition to the Wilson line
phases to give an expression for the effective 
potential, and gauge symmetry breaking is discussed.
%%%%
%We evaluate one-loop effective
%potential by taking vacuum expectation values for adjoint 
%scalar field in vectormultiplet into account in addition 
%to the Wilson line phases. 
%%%
In evaluating the potential, we employ 
the Scherk-Schwarz mechanism 
and introduce bare mass for gaugino in order to 
break supersymmetry.  
%%%
%We present the expression of the effective  
%potential and study the vacuum structure of the model for 
%the $SU(N)$ gauge group. 
%%%%%
We also obtain masses for the scalars, the adjoint  
scalar, and the component gauge field for the $S^1$ direction 
in case of the $SU(2)$ gauge group. We observe that large
supersymmetry breaking gives larger mass for the scalar.
This analysis is easily applied to the $M^4\times S^1/Z_2$ case.
\end{abstract}
\end{titlepage}
%%%%%%%%%%
%\tableofcontents
%%%%%%%%%%%%
\newpage
%%%%%%%%% INTRODUCTION %%%%%%%%
\section{Introduction}
Since the pioneering work by Hosotani \cite{hosotania}, gauge 
symmetry breaking
through the Wilson lines (Hosotani mechanism) has been an attractive 
mechanism in physics with extra dimensions. Namely, the mechanism is
expected to play the crucial role for the idea of the
gauge-Higgs unification \cite{gaugehiggs1}-\cite{hoso} and can 
provide a new framework for the grand unified theory. 
\par
%%%%%%%%%%%  
If extra dimensions are compactified on a certain topological manifold,
component gauge fields, which in fact behave like the adjoint Higgs 
scalars at low energies, can develop vacuum expectation values to induce 
dynamical gauge symmetry breaking. Gauge symmetry breaking patterns
have been studied extensively from various points of view in many
models, including supersymmetric gauge models \cite{models}. Gauge symmetry
breaking is usually studied by evaluating effective potential perturbatively 
for the Wilson line phases, which are related with the eigenvalues of 
the component gauge field for the compactified direction.
\par
%%%%%%%%%%%
One introduces the supersymmetric Yang-Mills theory in five dimensions
when one studies the scenario of the gauge-Higgs unification. 
In five dimensions the vector multiplet consists of 
the gauge field $(A_{\hmu})$, a real scalar $(\Sigma)$ and a Dirac 
spinor $(\lambda_D)$ \cite{sohnius}. The Dirac spinor 
is decomposed into two symplectic
Majorana spinors, $\lambda, \lambda^{\prime}$. Let us note that
one needs the real scalar $\Sigma$ in order to match the on-shell
degrees of freedom between the bosons and fermions in the supermultiplet.
\par
%%%%%%%%%%%
If one of the space coordinates is compactified on $S^1$, the component 
gauge field for the $S^1$ direction $A_y$ becomes a dynamical variable 
and its vacuum expectation 
values $\vev{A_y}$ cannot be gauged away, reflecting the topology of
$S^1$. Depending 
on the values of $\vev{A_y}$, the gauge symmetry is dynamically broken
down \cite{hosotania}. It should be noted here that, in addition to
$\vev{A_y}$, the vacuum expectation values of $\Sigma$, which is
the adjoint scalar field, are also order
parameters for gauge symmetry breaking. Even if one tries to
remove $\vev{A_y}$ by a singular gauge transformation,
such a gauge transformation changes boundary conditions of fields for
the $S^1$ direction. Therefore, it is impossible to remove both of 
the vacuum expectation values from 
the theory. Taking $\vev{\Sigma}$ into
account, though important, has been
overlooked in many papers studying the gauge-Higgs unification 
scenario in five dimensional supersymmetric gauge 
models \footnote{As we will see later,  
taking $\vev{\Sigma}=0$ from the beginning is justified, {\it a posteriori}, in
some case.}. There are two kinds of order parameter for gauge symmetry 
breaking in the theory, one is $\vev{A_y}$, which has a periodicity of
$2\pi$, reflecting the original five dimensional local gauge invariance 
and the other one is $\vev{\Sigma}$. In order to study the vacuum 
structure by evaluating the effective potential, one 
should take both $\vev{A_y}$ and $\vev{\Sigma}$ into account.
\par
%%%%%%%%%
In this paper we study the one-loop effective potential for 
the five-dimensional Yang-Mills theory defined on $M^4\times S^1$ by 
taking both $\vev{A_y}$ and $\vev{\Sigma}$ into account. 
To our best knowledge, this is the
first paper that studies the effective potential by taking account
of the two kinds of order parameter for gauge symmetry breaking. We
will give the expression for the
potential in one-loop approximation. We study 
the case of the $SU(N)$ gauge group and determine the vacuum expectation
values for $\vev{A_y}$ and $\vev{\Sigma}$ dynamically. 
We also evaluate masses for 
$A_y$ and $\Sigma$ for the $SU(2)$ gauge group, which 
are generated by quantum effects. We numerically obtain the masses with
respect to the change of the supersymmetry breaking parameters.
\par
%%%%%%%%%%%
\section{Effective potential of model}
We start with the five-dimensional Yang-Mills theory. The Lagrangian is given by
\begin{equation} 
{\cal L}=\mbox{tr}\left(
-\half F_{\hmu\hnu}F^{\hmu\hnu}+ 
D_{\hmu}\Sigma D^{\hmu}\Sigma + {\bar\lambda}_D i\Gamma^{\hmu}D_{\hmu}\lambda_D 
-g{\bar\lambda}_D[\Sigma,~\lambda_D]\right),
\end{equation}
where 
\begin{equation}
F_{\hmu\hnu}\equiv \del_{\hmu}A_{\hnu}
-\del_{\hnu}A_{\hmu}-ig[A_{\hmu},~A_{\hnu}],\quad
D_{\hmu}\phi\equiv \del_{\hmu}\phi -ig [A_{\hmu},~\phi], 
\quad \phi=\Sigma, \lambda_D.
\end{equation}
%%%%%%%%%%%
$\lambda_D$ is a Dirac spinor. $\hmu, \hnu$ run 
from $0$ to $4$, $x^{\mu}$ 
stands for the coordinates 
of the four-dimensional Minkowski
space-time, and $y$ denotes the coordinate of $S^1$. For a moment, we
consider the $SU(N)$ gauge group. We assume that all the fields satisfy
the periodic boundary conditions.  
\par
%%%%%%%%%%%%%
We evaluate the effective potential in one-loop approximation by
expanding fields around the vacuum expectation values 
\begin{equation}
A_{\hmu}=\vev{A_{\hmu}}\delta_{\hmu y} + {\bar A}_{\hmu},\qquad
\Sigma = \vev{\Sigma} +{\bar \Sigma}
\end{equation}
and by keeping the quadratic terms with respect to the fluctuations.
As noted in the introduction, one needs to take both $\vev{A_y}$ and
$\vev{\Sigma}$ into account. It is convenient to choose 
the gauge fixing term as
\begin{equation}
{\cal L}_{gf}=-{1\over \xi}tr
\Biggl(\del_{\mu}{\bar A}^{\mu}-\xi \biggl(D_y {\bar A}_y
-ig[\vev{\Sigma},~{\bar \Sigma}]\biggr) \Biggr)^2,
\label{fixing}
\end{equation}
%%%%%%%%%%%
where $\xi$ is the gauge parameter and 
\begin{equation}
D_y{\bar A}_y\equiv \del_y{\bar A}_y -ig [\vev{A_y},~{\bar A}_y].
\end{equation}
The 'tHooft-Feynman gauge $\xi=1$ makes the expression 
simple, as shown in the appendix, where the detailed 
expressions for the quadratic terms are given.
\par
%%%%%%%%
There arises the tree-level 
potential (\ref{tree}), which
is given by the commutator between $A_y$ and $\Sigma$. 
By utilizing the global gauge degrees of freedom, $\vev{A_y}$, for
example, can be diagonalized. It is natural to
expect that the vacuum configuration is given by the one
satisfying the flat direction,
\begin{equation}
[\vev{A_y},~\vev{\Sigma}]=0.
\end{equation}
This means that both $\vev{A_y}$ and $\vev{\Sigma}$ take diagonal forms 
simultaneously. Thus, we parameterize
them as \footnote{Let us note that the combination $gL\vev{A_y(\Sigma)}$ is a
dimensionless quantity.}
\begin{eqnarray}
gL\vev{A_y}&=&{\rm diag}(\theta_1, \theta_2, \cdots, \theta_N)
\quad \mbox{with}\quad \sum_{i=1}^{N}\theta_i=0.\label{vevuno}\\
\vev{\Sigma}&=&{\rm diag}(v_1, v_2, \cdots, v_N),
\quad \mbox{with}\quad \sum_{i=1}^{N}v_i=0. \label{vevdue}
\end{eqnarray}
\par
%%%%%%%%%%%%%
It is important to note that one can redefine the field in
such a way that $\vev{A_y}$ is removed by a singular gauge
transformation, but, accordingly, the 
boundary condition of field is 
twisted by an amount of the vacuum expectation values. Thus, one
cannot remove both $\theta_i$'s and
$v_i$'s from the theory, so that we have two
kinds of the order parameters for the 
gauge symmetry breaking, and $\theta_i$'s are related with the Wilson line
phase
\begin{equation}
W={\cal P}\mbox{exp}\left(\oint_{S^1}dy\vev{A_y}\right)
=\mbox{diag}(\e^{i\theta_1}, \e^{i\theta_2}, \cdots, \e^{i\theta_N})
\label{wilson}
\end{equation}
and are modules of $2\pi$. This reflects the fact that
$A_y$ is a part of the five dimensional gauge potential $A_{\hmu}$ and
is subject to the five dimensional local gauge transformation. On the
other hand, the vacuum expectation 
values for $\Sigma$ does not have such the periodicity, and $\Sigma$ is 
just a
real scalar field belonging to the adjoint representation under the
gauge group.
\par 
%%%%%%%%%%
Inserting the vacuum expectation 
values (\ref{vevuno}) and (\ref{vevdue}) 
into Eqs.(\ref{gauge})$\sim$(\ref{dirac}) in the appendix, we arrive at 
Eqs.(\ref{vevsigma})$\sim$(\ref{vevdirac}). 
Then, the effective potential in one-loop approximation is obtained as
%%%%%
\begin{equation}
V_{eff}=\left(\sum_{i=A_{\mu}, A_{y}, \Sigma, \lambda_D}
N_{deg}^i\times (-1)^{fermion}\right)\times I,
\end{equation}
where
\begin{equation}
I\equiv \half \sum_{i, j=1}^{N}\sum_{n=-\infty}^{\infty}
{1\over L}\int {{d^4p_E}\over {(2\pi)^4}}ln 
\Biggl[p_E^2 +\Biggl({{2\pi}\over L}\Biggr)^2
\Biggl(n-{{\theta_i -\theta_j}\over {2\pi}}\Biggr)^2 + g^2(v_i - v_j)^2\Biggr]
\label{integration}
\end{equation}
and $(-1)^{fermion}=-1$ for fermions is due to fermi statistics.
$N_{deg}^i$ stands for the on-shell degrees of freedom such as
$(4-2)$ for $A_{\mu}$, $1$ for each $A_y$ and $\Sigma$, and $4$ for $\lambda_D$.
Here we have made the Wick rotation for the four-dimensional momentum.
\par
%%%%%%%%%%
We immediately observe that the effective potential $V_{eff}$ vanishes
due to supersymmetry,
\begin{equation}
\left(\sum_{i=A_{\mu}, A_{y}, \Sigma, \lambda_D} 
N_{deg}^i\times (-1)^{fermion}\right)=0.
\end{equation}
In order to have nonvanishing effective
potential, one needs to break supersymmetry somehow. One of the simple
ways to break supersymmetry is to resort to the 
Scherk-Schwarz mechanism \cite{ss}. In the mechanism the boundary 
conditions of the gauge fermions for the $S^1$ direction are twisted,
\begin{equation}
\left(
\begin{array}{c}
\lambda\\
\lambda'
\end{array}\right)
(x^{\mu}, y+L)=\e^{i\beta\sigma_3}
\left(
\begin{array}{c}
\lambda\\
\lambda'
\end{array}\right)
(x^{\mu}, y).
\end{equation}
The nontrivial phase $\beta$ shifts the Kaluza-Klein modes and modifies
the momentum for the $S^1$ direction as
\begin{equation}
p_y\equiv {{2\pi}\over L}\left(n +{{\theta_i-\theta_j}\over {2\pi}}\right)
\rightarrow {{2\pi}\over L}
\left(n + {{\theta_i-\theta_j-\beta\sigma_3}\over {2\pi}}\right). 
\end{equation} 
Moreover, it is also possible to break supersymmetry by introducing the
gauge invariant bare mass term $Mtr({\bar \lambda}_D\lambda_D)$
for $\lambda_D$ \cite{takenaga}. In this case, we have the modification given by
\begin{equation}
g^2(v_i - v_j)^2 \rightarrow g^2(v_i - v_j)^2 + M^2.
\end{equation} 
Supersymmetry is explicitly broken for both cases. 
\par
%%%%%%%%
Following the standard prescription \cite{pomarol}, we obtain that for the
bosonic fields, 
%%%
%up to $\theta_i$-independent terms,  
%%%
\begin{eqnarray}
I^{b}&\equiv &{{-2}\over {(2\pi)^{5\over 2}}}
\sum_{i,j=1}^{N}\sum_{n=-\infty,\neq 0}^{\infty}
\Biggl({{g^2(v_i-v_j)^2}\over {n^2L^2}}\Biggr)^{5\over 4}
K_{5\over 2}\Biggl(\sqrt{g^2(v_i-v_j)^2n^2L^2}\Biggr)~~
\e^{-in(\theta_i-\theta_j)} \nonumber\\
&=&{{-3}\over {4\pi^2}}\sum_{i, j=1}^{N}\sum_{n=1}^{\infty}
{1\over {n^5L^5}}
\Bigl[1+gv_{ij}nL+{{(gv_{ij}nL)^2}\over 3}\Bigr]\e^{-gv_{ij}nL}
\cos[n(\theta_i -\theta_j)],
\label{eqboson}
\end{eqnarray}
where $v_{ij}\equiv \abs{v_i - v_j}$ and we have used the formula
for the modified Bessel function,
\begin{equation}
K_{5\over 2}(y)=\left({\pi\over {2y}}\right)^{\half}
\left(1+{3\over y}+{3\over{y^2}}\right)
\e^{-y}.
\label{suppress}
\end{equation}
On the other hand, for the fermionic field, taking account of the supersymmetry
breaking discussed above, we obtain that 
\begin{eqnarray}
I^{f}&\equiv&{{-2}\over {2(2\pi)^{5\over 2}}}
\sum_{i,j=1}^{N}\sum_{n=-\infty, \neq 0}^{\infty}
\Biggl[{{g^2(v_i-v_j)^2+M^2}\over {n^2L^2}}\Biggr]^{5\over 4}\nonumber\\
&\times &K_{5\over 2}\Biggl(\sqrt{(g^2(v_i-v_j)^2+M^2)n^2L^2}\Biggr)
\times \half\Biggl(\e^{-in(\theta_i-\theta_j-\beta)}+\e^{-in(\theta_i-\theta_j+\beta)}   
\Biggr)\nonumber \\
&=&{{-3}\over {4\pi^2}}\sum_{i, j=1}^{N}\sum_{n=1}^{\infty}
{1\over {n^5L^5}}
\Bigl[1+nL\sqrt{g^2v_{ij}^2 + M^2} +
{{g^2v_{ij}^2 + M^2}\over 3}n^2L^2 \Bigr]\nonumber\\
&\times & \e^{-\sqrt{M^2+g^2v_{ij}^2}nL}
\cos[n(\theta_i -\theta_j-\beta)].
\label{eqfermi}
\end{eqnarray}
%%%
%We have also ignored the irrelevant constants in Eq.(\ref{eqfermi}). 
%%%%
Let us note that $I^b, I^f$ are even function of $v_i-v_j$, so that it is
enough to consider the case $v_i-v_j \geq 0$.
\par
%%%%%%%%
If supersymmetry is broken by the Scherk-Schwarz mechanism 
alone, the divergent terms that depend on the order parameters 
are absent in Eqs.(\ref{eqboson}) and (\ref{eqfermi}). In this case, the effective 
potential does not suffer from ultraviolet effects, reflecting 
the supersoft property of the Scherk-Schwarz mechanism. 
%the fact
%that the Wilson line phases are global, while ultraviolet effects are local. 
%%
Here we have also introduced the supersymmetry breaking bare 
mass, by which there appear the order parameter $v_i$-dependent divergent terms.
%%%%%
%There are $v_i$-dependent divergent terms in
%Eq.(\ref{integration}), which have the form of $M^2(v_i-v_j)^2\Lambda$. 
%%%%%
We have made the regularization of Eqs.(\ref{eqboson}) and (\ref{eqfermi}) 
at $L\rightarrow \infty$ (implying $\vev{A_y}=0$), so that the terms 
vanish, which formally corresponds to subtracting $n=0$ mode in the summation. 
%%%
%We also notice 
%that $\vev{\Sigma}$ gives the mass 
%terms for ${\bar A}_{\hmu}, {\bar \Sigma}$ and $\lambda_D$, as seen 
%from Eqs.(\ref{vevsigma}) $\sim$ (\ref{vevdirac}) in the 
%appendix. 
%%%%
\par
%%%%%%%%%%%%
Collecting the contributions from the boson and fermion,  the effective 
potential is given by
\begin{eqnarray}
V_{eff}(v_{ij}, \theta_i)&=&4I^{b}+(-1)\times 4I^{f} \nonumber\\
&=&-4\biggl({3\over{4\pi^2}}\biggr) \sum_{n=1}^{\infty}{1\over {n^5L^5}}
\left[B(v_{ij}, n)-F(v_{ij}, n, M) \cos(n\beta)\right]
\nonumber\\
&\times & \sum_{1\leq i < j \leq N}
2\cos[n(\theta_i-\theta_j)].
\label{effpot}
\end{eqnarray}
where $B(v_{ij}, n)$ comes from the bosonic contributions in
the vector multiplet and is given by
\begin{equation}
B(v_{ij}, n)\equiv 
\left(1+gv_{ij}nL+{{g^2v_{ij}^2n^2L^2}\over 3}\right)\e^{-gv_{ij}nL},
\end{equation}
while the fermionic contribution $F(v_{ij}, M, n, L)$ is 
\begin{equation}
F(v_{ij}, n, M)\equiv 
\left(1+nL\sqrt{g^2v_{ij}^2 + M^2}
+{{g^2v_{ij}^2 + M^2}\over 3}n^2L^2\right)\e^{-\sqrt{M^2+g^2v_{ij}^2}nL}.
\end{equation}
We observe that supersymmetry is broken by either the Scherk-Schwarz
mechanism or the bare mass $M$ for $\lambda_D$ to yield 
the nonvanishing effective
potential. Supersymmetry restores by taking  
$\beta=2\pi{\bf Z}~({\bf Z}\in \mbox{integer})$ and the limit 
$M \rightarrow 0$ simultaneously. As for 
the parameter $\beta$, it is enough to consider the
region of $0\leq \beta \leq \pi$.
\par
%%%%%%%%%%
We also note here that the 
effective potential (\ref{effpot}) shares many
similarities with the potential obtained in finite temperature field
theory. The particles in the theory become massive due to
$\vev{\Sigma}$ (and $M$), so that, as is well known, particles with
smaller wavelengths than
the inverse temperature $(\sim L)$ have the 
Boltzmann (exponentially) suppressed distribution in the system.  
It has been known that the Boltzmann-like suppression 
factor is important for the gauge symmetry breaking through the Hosotani
mechanism \cite{yamashita}\cite{takenaga}.
%%%%%%%%%%%%%%%%%%%%
\section{Vacuum structure and mass terms for $A_y$ and $\Sigma$}
Let us study the vacuum structure of the model. By minimizing the
effective potential (\ref{effpot}) with respect to $\theta_i$'s and
$v_i$'s, those order parameters are dynamically determined. 
It is important to note that for any values 
of $v_{ij}, n, M$ and $\beta$, we have  
\begin{equation}
B(v_{ij}, n)-F(v_{ij}, n, M) \cos(n\beta)\geq 0,
\end{equation}
where the equality holds if and only if $\beta=2\pi{\bf Z}$ and $M=0$
are satisfied simultaneously. Then, we see that
$\theta_i=\theta_j$
gives the lowest energy configuration for fixed values of $v_{ij}$
because of the overall minus sign in the effective potential.
Taking $\sum_{i=1}^{N}\theta_i=0$ into account, we obtain that
\begin{equation}
\theta_i={{2\pi k}\over N}~~(\mbox{mod}~~2\pi),\qquad k=0, 1, \cdots, N-1,
\label{thetavev}
\end{equation}
for which the Wilson line (\ref{wilson}) gives 
the center  of $SU(N)$. The configuration (\ref{thetavev}) does not
break the $SU(N)$ gauge symmetry. Now we  
observe that $v_{ij}\rightarrow 0$ gives the lowest
energy configuration for the given values 
of $\theta_i$'s obtained in Eq.(\ref{thetavev}). Thus, we have
$v_i=0~(i=1, 2, \cdots, N)$ because $\sum_{i=1}^{N}v_i=0$. Therefore,
the vacuum configuration of the model is
dynamically determined as
\begin{equation}
(\theta_i, v_i)=({{2\pi k}\over N},~~0),~~(i=1, 2, \cdots, N-1),
\label{vacuum}
\end{equation}
so that the $SU(N)$ gauge symmetry is not broken.
\par
%%%%%%%%%%%%%%
Let us next consider the case of the $SU(2)$ gauge group and study 
the masses for $A_y$ and $\Sigma$. Denoting $v_1=-v_2\equiv v$ and
$\theta_1=-\theta_2=\theta$, the vacuum configuration in this case is
given by 
\begin{equation}
(\theta,~v)=(0~(\mbox{mod}~\pi), 0),
\label{vevmin}
\end{equation}
for which the $SU(2)$ gauge symmetry is not broken. In Fig. $1$, we
depict the behavior of the effective potential for the 
parameter $(\beta/2\pi,~ML)=(0.1,~1.0)$.
\begin{figure}
\centering
\leavevmode
\includegraphics[width=9.cm]{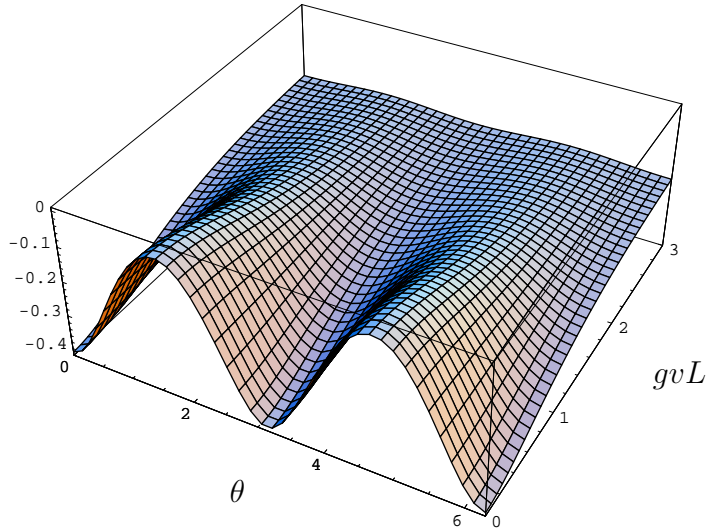}
\begin{picture}(0,0)
\put(-175,15){$\theta$}
\put(-15,60){$gvL$}
\end{picture}
\caption{The behavior of the effective potential (\ref{effpot}) 
for $(\beta/2\pi, ML)=(0.1, 1.0)$. The gauge group is $SU(2)$.}
\end{figure}
\par
%%%%%%%%%%%%%%
%\section{Mass terms for $A_y$ and $\Sigma$}
By evaluating the tree-level potential (\ref{tree}) and the 
second derivative of the effective potential with respect
to $\theta$ and $v$ at the
minimum (\ref{vevmin}), we obtain the masses for $A_y$ and $\Sigma$ in one-loop
approximation. Contrary to the modes ${\bar A}_y^{a=3}$ and
${\bar\Sigma}^{a=3}$, there arises the mass terms 
for ${\bar A}_y^{a=1,2}$ and ${\bar\Sigma}^{a=1, 2}$ in the background
(\ref{vevuno}) and (\ref{vevdue}) at the tree-level from the commutator 
between $A_y$ and $\Sigma$. It is calculated in the basis of 
$({\bar A}_y^{1}, {\bar A}_y^{2},{\bar\Sigma}^{1},{\bar\Sigma}^{2})^T$ as
\begin{equation}
{\cal M}_{tree}^2=\left(
\begin{array}{cccc}
4g^2v^2 & 0 & -4\theta v g/L & 0 \\
0 & 4g^2v^2 & 0 & -4\theta v g/L \\
-4\theta v g/L & 0 & 4\theta^2/L^2 & 0 \\
0 & -4\theta v g/L & 0 & 4\theta^2/L^2 
\end{array}
\right).
\end{equation}
The eigenvalues of the matrix are given by 
\begin{equation}
0~(\mbox{degeneracy}=2,)\quad 
\mbox{and}\quad 4(g^2v^2+\theta^2/L^2)~(\mbox{degeneracy}=2).
\end{equation}
The two massless modes are the Nambu-Goldstone bosons absorbed by the
charged massive gauge boson, and the rest corresponds to
the charged massive state under the survived $U(1)$ gauge symmetry after 
the breakdown of the $SU(2)$ gauge symmetry.
\par
%%%%%%%%%%
The zero modes ${\bar A}_y^{a=3}$ and ${\bar\Sigma}^{a=3}$ become
massive at the one-loop level. The masses are evaluated by the second
derivative of the effective potential evaluated at 
the vacuum configuration (\ref{vevmin})
\begin{equation}
{\cal M}_{A_y, \Sigma}^2=\left(
\begin{array}{cc}
{{\del^2 V_{eff}}\over {\del\theta^2}} & {{\del^2 V_{eff}}\over{\del\theta \del v}} 
\\[0.3cm]
{{\del^2 V_{eff}}\over{\del\theta \del v}}&{{\del^2 V_{eff}}\over {\del v^2}} 
\end{array}
\right)_{vac}.
\end{equation}
For the vacuum configuration, the off-diagonal elements vanish, so that the 
masses for ${\bar A}_y^{a=3}, {\bar\Sigma}^{a=3}$ are given by
$\frac{(gL)^2}{4}{{\del^2 V_{eff}}\over {\del\theta^2}}, 
\frac14{{\del^2 V_{eff}}
\over {\del v^2}}$, respectively. Thus, we obtain 
that \footnote{The squared masses for the zero modes are
proportional to the number
of colors $N$ if one considers the $SU(N)$ gauge group. It may be
natural to impose $g^2N < O(1)$.} 
\begin{eqnarray}
m_{\Sigma}^2&=&\left({g_4\over L}\right)^2{{2\times 1}\over
\pi^2}\sum_{n=1}^{\infty}
{1\over n^3}\left[ 1-(1+nML)\e^{-nML}\cos(n\beta)\right]\geq 0,
\label{massuno}\\
m_{A_y}^2&=&\left({g_4\over L}\right)^2{{2\times 3}\over \pi^2}
\sum_{n=1}^{\infty}
{1\over n^3}\left[ 1-(1+nML+{{(nML)^2}\over 3})
\e^{-nML}\cos(n\beta)\right]\geq 0,
\label{massdue}
\end{eqnarray}
where we have defined the four dimensional 
gauge coupling $g_4\equiv g/\sqrt{L}$.
The equality holds if and only if $\beta=2\pi{\bf Z}$ and $M=0$ are
satisfied simultaneously, for which supersymmetry
restores. 
\par
%%%%%%%%
Let us first discuss the mass scale of $m_{A_y, \Sigma}$.
The mass scale of the generated mass is roughly estimated as 
\begin{equation}
m_{\Sigma},~m_{A_y}\simeq cg_4\times 
\left\{\begin{array}{lll}
1/L     &\mbox{for} & ML\geq  O(1), \\[0.3cm]
%%%%
%\beta/L      &\mbox{for} & ML << 1,~\beta\sim O(1),     \\[0.3cm]
%M~~\mbox{or}~~\beta/L&\mbox{for} & ML << 1,~\beta <<1,          \\[0.3cm] 
%%%%
\mbox{max}~(M,~\beta/L)&\mbox{for} & ML << 1,          \\[0.3cm]
\end{array}
\right.
\label{higgsmass}
\end{equation}
where $c$ is a numerical constant of order $O(1)$ and
$\mbox{max}~(M,~\beta/L)$ stands for the larger scale among $M, \beta/L$. 
In order for ${\bar A}_y, {\bar\Sigma}$ to become 
massive, one needs the breaking of both supersymmetry and the 
five dimensional local gauge
invariance simultaneously. The former scale is given by the 
scale, $\mbox{max}~(M,~\beta/L)$, while the latter is given 
by the compactification scale $L^{-1}$.
The mass scale should be the one at which both breaking are occurred.
In fact, as observed in Eq.(\ref{higgsmass})
the mass scale of $m_{A_y, \Sigma}$ is the order of the scale at which 
the breaking is realized.
\par
%%%%%%%%
As the bare mass $M$ becomes larger and larger, the contribution from
the fermion $\lambda_D$ to the effective potential is suppressed more and more
thanks to the Boltzmann factor in Eq.(\ref{suppress}), and what is left is the
contribution from the boson alone. The parameter $\beta$ has no
effect on the size of the mass in the heavy bare mass limit. The values of 
$\beta$ affects to the size of the mass for $ML \leq  O(1)$. As shown in 
Table $1$, we see that $\beta/2\pi=0.5$, which corresponds to 
the antiperiodic boundary condition
for the fermions and is the ``maximal'' breaking of supersymmetry,
significantly enhances the masses. This is because 
for $\beta/2\pi=0.5$, the second term 
in Eqs.(\ref{massuno}) and (\ref{massdue}) becomes negative, and then,
the contribution to the effective potential is additive to make the masses larger.
It is important to study the supersymmetry breaking effect 
on the magnitude of the mass for the Higgs scalar in the scenario of the
gauge-Higgs unification \cite{next}.
%%%%%%%%%%%
\begin{table}
\begin{center}
\begin{tabular}{l|l|l}
\hline
$\beta/2\pi$ & ${\bar m}_{A_y}(ML=1.0)$ & ${\bar m}_{\Sigma}(ML=1.0)$ \\ \hline
$0.0$     & 0.503404     &  0.637422  \\ \hline
$0.0001$  & 0.503404    &  0.637422  \\ \hline
$0.001$  & 0.50343    &  0.637438  \\ \hline
$0.01$   & 0.506036    &  0.63903   \\ \hline
$0.1$    & 0.701635    &  0.771337  \\ \hline
$0.5$     & 1.4134    &  1.37596 \\ \hline\hline
$\beta/2\pi$ & ${\bar m}_{A_y}(ML=0.1)$ & ${\bar m}_{\Sigma}(ML=0.1)$ \\ \hline
$0.0$  & 0.0796115    &  0.119074 \\ \hline
$0.0001$  & 0.079616    &  0.119076  \\ \hline
$0.001$  & 0.0800608    &  0.119347  \\ \hline
$0.01$  & 0.115059    &   0.143266 \\ \hline
$0.1$  & 0.623837   &   0.62522 \\ \hline
$0.5$  & 1.44998   &    1.44924 \\ \hline
\end{tabular}
\end{center}
\caption{The masses for $A_y$ and $\Sigma$ with respect to the values of
 $\beta/2\pi$ for fixed values of $ML=1.0$ and $0.1$. 
${\bar m}_{A_y}\equiv {{m_{A_y}}\over 
{\left({g_4\over L}\right){{\sqrt{6}}\over \pi}}},~~{\bar m}_{\Sigma}\equiv 
{{m_{\Sigma}}\over {\left({g_4\over L}\right){{\sqrt{2}}\over \pi}}}$.}
\end{table}
%%%%%
%We also observe that even though $M << L^{-1}$, the original 
%supersymmetry protects  the mass term 
%for ${\bar A}_y^{a=3} ({\bar\Sigma}^{a=3})$
%against the large correction of order $O(L^{-1})$. 
%This is an expected thing, but here we have demonstrated it explicitly 
%in the simple model. 
%
%Let us
%also note that the nontrivial phase arising from the Scherk-Schwarz
%mechanism also reduces the size of the masses if $\beta<<1$ for the case
%of $ML<< 1$.  
\par
%%%%%%%%%%%
\section{Conclusions and Discussions}
We have studied the supersymmetric Yang-Mills theory on $M^4\times
S^1$ by taking the two kinds of order parameter for the gauge symmetry
breaking into account. One is the component gauge 
field $A_y$ for the $S^1$ direction, and
the other one is the real scalar field $\Sigma$. The latter has been
overlooked in the past. We have evaluated the 
effective potential for the order parameters in one-loop approximation. 
In the calculation we have employed the Scherk-Schwarz
mechanism and have introduced the bare mass for
$\lambda_D$ in order to break supersymmetry to yield 
the nonvanishing effective potential (\ref{effpot}).   
\par
%%%%%%%
The effect of the vacuum expectation values $\Sigma$ and the bare mass
$M$ appears as the Boltzmann suppression factor in the effective 
potential. This can be understood from the 
similarity of the potential with the one obtained
in finite temperature field theory as explained in the text. 
\par
%%%%%%%%%%%
We have first studied the effective potential for the $SU(N)$ gauge 
group, and, by minimizing the potential, we have obtained the vacuum 
configuration (\ref{vacuum}), for which the gauge symmetry is not broken. 
We have also evaluated the masses 
for ${\bar A}_y^{a=3}$ and ${\bar\Sigma}^{a=3}$ 
by the second derivative
of the effective potential at the vacuum configuration for the case 
of $SU(2)$. These masses
are generated by the quantum effect due to the breaking of both
supersymmetry and the five dimensional local gauge 
invariance. Hence, the mass scale should be the order of the
scale at
which both symmetries are violated, as evaluated in Eq.(\ref{higgsmass}).
\par
%%%%%%%%%%
The suppression factor arising from $\vev{\Sigma}$ in the effective
potential 
also appears for the case of orbifold compactification such 
as $S^1/Z_2$ \cite{habab}\cite{haba}\cite{toshi}. This point has 
been overlooked in the past. As shown in the example, however, $\vev{\Sigma}$
takes the values of zero dynamically, and it is also the case for the
orbifold compactification. And even if we introduce matter into the 
theory, we
expect $\vev{\Sigma}=0$ at the minimum of the effective 
potential. Therefore, it is justified 
to consider the Wilson 
line phases alone, {\it a posteriori}, in evaluating the effective
potential. It is 
more important to study 
the effect of the vacuum expectation values of squark field
$\vev{\phi_i}$ in hypermultiplet and is interesting to find the case, where
$\vev{\phi_i}$ take nontrivial values. 
%%%%%%%%%%
%In our examples, the vacuum expectation values $\vev{\Sigma}$ is
%dynamically determined to be zero. And even if we introduce matter fields
%belonging to the adjoint or fundamental representation under the gauge
%group, it is likely that the vacuum expectation values, denoted 
%by $v_i$, arising from the
%squark field vanish at the vacuum configuration. This is because for
%fixed values of $v_i$, the effective potential is expected to take 
%$negative values for
%the configuration that gives the lowest energy. Hence, the vanishing
%$v_i$ makes the potential energy more deeper toward negative 
%direction \footnote{For positive values of the potential, the potential
%has the runaway type behavior.}. Hence, taking  
%only the Wilson line phases seems to be justified {\it a posteriori} when we
%study the effective potential. It may be interesting, however, if there
%exist some models which yield nontrivial values of $v_i$ as 
%the vacuum configuration.
%%%%%%%%
\vspace{3cm}
%%%%%%%%%%%%%
\begin{center}
{\bf Acknowledgements}
\end{center}  
N.H. is supported in part by the Grant-in-Aid for Science
Research, Ministry of Education, Science and Culture, Japan,
No.~16540258, 16028214, 14740164. K.T. would thank the colleagues 
in Osaka University and the professor Y. Hosotani for 
valuable discussions, and he is supported by the $21$st Century 
COE Program at Osaka University. T.Y. would like to thank H.Kogetsu for
useful discussion, and the Japan Society
for the Promotion of Science for financial support. 
\vskip 2cm
%%%%%%%%%%%%%%
\begin{center}
{\bf Appendix}
\end{center}
Equipped with the gauge fixing term (\ref{fixing}) given in the 
text, the quadratic terms with
respect to the fluctuations are given by
\begin{equation}
{\cal L}_{eff}={\cal L}_{tree}+{\cal L}_{A_{\mu}}+{\cal L}_{A_{y}}+{\cal L}_{\Sigma}
+{\cal L}_{\lambda_D}+{\cal L}_{others}
\end{equation}
where 
\begin{eqnarray}
{\cal L}_{tree}&=&g^2tr[\vev{A_y},~\vev{\Sigma}]^2,\label{tree}\\
{\cal L}_{A_{\mu}}&=& -\half tr(F_{\mu\nu}F^{\mu\nu})- 
g^2tr[{\bar A}_{\mu},~\vev{\Sigma}][{\bar A}^{\mu},~\vev{\Sigma}]\nonumber\\
&+&tr(D_y{\bar A}_{\mu})^2-{1\over \xi}tr(\del_{\mu}{\bar A}^{\mu})^2,\label{gauge}\\
{\cal L}_{A_{y}}&=&tr(\del_{\mu}{\bar A}_y)^2-\xi tr(D_y{\bar A}_y)^2
+g^2tr[{\bar A}_y,~\vev{\Sigma}]^2,\label{gaugey}\\
{\cal L}_{\Sigma}&=&tr(D_{\hmu}{\bar\Sigma})^2
+\xi g^2 tr[\vev{\Sigma},~{\bar\Sigma}]^2,\label{fsigma}\\
{\cal L}_{\lambda_D}&=&tr({\bar\lambda_D}i\Gamma^{\hmu}D_{\hmu}\lambda_D 
-g{\bar\lambda}_D[\vev{\Sigma},~\lambda_D]),\\
{\cal L}_{others}&=&2ig(\xi -1)tr(D_y{\bar A}_y[\vev{\Sigma},~{\bar\Sigma})]\nonumber\\
&-&2ig~tr(\del_{\mu}{\bar\Sigma}[{\bar A}^{\mu}, \vev{\Sigma}])
-2ig~tr(\del_{\mu}{\bar A}^{\mu}[\vev{\Sigma}, \bar\Sigma])\label{dirac}\nonumber\\
&-&2g^2tr[\vev{\Sigma},~\vev{A_y}][{\bar\Sigma}, {\bar A}_y],
\end{eqnarray}
where the covariant derivative is defined in the background field
$\vev{A_y}$ as,
\begin{equation}
D_{\hmu}\phi \equiv \del_{\hmu}\phi-ig [\vev{A_{\hmu}}\delta_{\hmu y},~\phi],
\qquad \phi\equiv {\bar\Sigma},~{\bar A}_{\mu},~{\bar A}_y,~\lambda_D.
\end{equation}
We see that $\xi=1$ gives a simple expression. 
For the background fields defined by (\ref{vevuno}) and (\ref{vevdue}), we obtain that
\begin{eqnarray}
{\cal L}_{\Sigma}&=&
-tr{\bar\Sigma}\biggl(\del_{\mu}\del^{\mu}
-(\del_y -{i\over L}(\theta_i-\theta_j))^2 +\xi
g^2(v_i-v_j)^2\biggr){\bar\Sigma},\label{vevsigma}\\
{\cal L}_{A_{\mu}}&=&
tr{\bar A}_{\mu}\biggl(
\eta^{\mu\nu}[\del_{\rho}\del^{\rho}+ 
g^2(v_i - v_j)^2 -(\del_y-{i\over L}(\theta_i -\theta_j))^2 ]
-(1-{1\over \xi})\del^{\mu}\del^{\nu}\biggr){\bar A}_{\nu},
\label{vevgauge}\\
{\cal L}_{A_y}&=&
-tr{\bar A}_y
\biggl(\del_{\mu}\del^{\mu}- 
\xi(\del_y -{i\over L}(\theta_i -\theta_j))^2 + g^2(v_i -v_j)^2
\biggr){\bar A}_y,\label{vevgaugey}\\
{\cal L}_{\lambda_D}&=&
tr{\bar \lambda}_D\left(i\Gamma^{\hmu}\del_{\hmu}+\Gamma^y
{1\over L}(\theta_i-\theta_j)+g(v_i-v_j)\right)\lambda_D,\label{vevdirac}
\end{eqnarray}
where $\xi=1$ is understood. Let us note that ${\cal L}_{others}$
vanishes for the parameterization of the vacuum expectation 
values (\ref{vevuno}) and (\ref{vevdue}) in the text. 
%%%%%%%%%%%%%
\newpage
%%%%%%%%%%%%% BIBLIOGRAPHY %%%%%%%%%%%%%%%%%%%%

%%%%%%%%%%%%%%%%%%%%%%%%%%%%
\end{document}